# A Smart Intelligent Way of Video Authentication Using Classification and Decomposition of Watermarking Methods

T.Srinivasa Rao[1], Dr.Rajasekhar R Kurra[2]

[1](InformationTechnology,UshaRama College of Engineering & Technology, India)
[2](Principal, Sri Prakash College of Engineering & Technology, India)

**ABSTRACT :** Video Watermarking serves as a new technology mainly used to provide security to the illegal distribution of digital video over the web. The purpose of any video watermarking scheme is to embed extra information into video in such a way that must be perceptually undetectable while still holding enough information in order to extract the watermark beginning with the resultant video. Information which is embedded within the original image is a "Digital Watermark", which could be visible or invisible. To improved more security, embedding and extraction Watermark process should be complex against attackers. Recent research indicates SVD (Singular Value Decomposition) algorithms are employed owing to their simple scheme with mathematical function. In this proposed work an advanced SVD transformation algorithm is used for embedding and extraction process. Experimental results show proposed watermarking process is more secured than existing SVD approach.

**Keywords:** *Embedding, I m a g e e n c r y p t i o n , , S V D , W a t e r m a r k*

## I. INTRODUCTION

Watermarking has been considered to make a promising solution that will protect the copyright of multimedia data through Trans coding, because the embedded message is often found in the data. However, today, there isn't any evidence that watermarking techniques can achieve the last word target to retrieve the right owner information seen from the received data in fact kinds of content-preserving manipulations. As a consequence, watermarks can only be embedded within a limited space within the multimedia data. There is certainly always a biased advantage for the attacker whose target is simply to get rid of the watermarks by exploiting various manipulations within the finite watermarking embedding space.

Using the rapid increase of internet technologies in addition to digital multimedia processing, a large amount of data is easily accessible to everyone today. Therefore, various authentication schemes have been already proposed for verifying the authenticity of one's image, video or text content. The authentication techniques are basically classified as: digital watermark based and digital signature based schemes. A digital signature based technique treated either an encrypted or a signed hash value of image contents or image characteristics [1]. This digital signature scheme maintains its own drawback is that; it might detect the modification files, but cannot find the regions where the image has actually been modified [2]. To solve the issue of locating the region of modification, digital watermarking techniques have been proposed by most researchers [3]. Digital watermarking is a technique which involves two steps: (i) an algorithm to embed small authentication information called watermark content toward the host content. (ii) an algorithm to retrieve or extract the embedded watermark with less distortion. Watermarking techniques can possibly be broadly categorized into two groups: spatial domain methods and transform domain methods. The spatial domain methods embed by modifying directly on the pixels associated with an image [4]. The transform domain method involves modifying the transform domain coefficients. Within this paper, we target the authentication of video content by embedding watermark video into your cover video, effectively making our approach robust against possible attacks. There are numerous ways to insert watermark data directly into video. The basic way involves for the video for being sequence of still images or frames, after which embeds each watermark frame into each cover frame independently. Here, we proposed a strong and imperceptible video watermarking algorithm combines two powerful mathematical transforms: Discrete Wavelet Transform, and to discover the Singular Value Decomposition (SVD).In addition to this, as a way to increase the grade of authentication, we also added the finger print of one's owner with the embedding level and at the receiver end , precisely the same will certainly be compared with the original fingerprint[2].

Properties of Digital Video Watermark
For digital watermarking of video, the various characteristics of this very watermarking are provided with below.
• Invisibility: The digital watermark we simply embed should be unseen by your eyes. To make sure that attacker does not aware the presence of watermark.
• Robustness: robustness is defined as the Attack that should be working on watermarked video and analyze just how it shows the resistant to various method of attack. a video watermark is truly robust then it can say that it having more resistant power. High robustness preserves the nature of video.
• Perceptible: A digital watermark is known as perceptible if the presence of your mark is noticeable. Getting Imperceptibility is great task for researcher.

• Capacity: capacity designates period of the embedded message into digital video.
• Fidelity: It is actually the similarity along at the point at which the watermarked content is provided towards the customer that count weather video handed to the public is degraded or not. A watermark is said to be high fidelity if degradation it causes is extremely difficult for a viewer to see.
• Computational Cost: it refers to the cost or time necessary for embedding and extracting the watermark from the digital video. For better working digital video watermarking scheme computational cost should be minimized.
• Interoperability: it refers, the watermark should carry on video including the compression and decompression operations are performed regarding that video.
• Blind/informed detection: among the Informed watermarking schemes the detector requires access to the unwatermarked original video. In Blind watermarking Detectors tend not to require any original information.
• False positive rate: An inaccurate positive refers detection of watermark information given by a digital media which can include video, image that really does not actually contain that watermark.

As digital video-based application technologies grow, which can include Internet video, wireless video, Video phones, and video conferencing, the challenge of illegal manipulation, copying, distribution and piracy of digital video rises increasingly more. The challenge with this paper research effort is to unravel the authentication problem and embed the watermark in such a manner that it could not be removed or damaged that are caused by the video utilizing the proposed algorithm of random frame selection through encryption key.

The watermark is embedded through these selected frames. Encryption key used is decided via the owner of the recording. And the random frames are selected by applying the functions generated using this authentication key. These functions are designed in a way that the avoiding the selection/clustering of frames in a chunk. Versus the clustering of frames, the frames are selected uniformly from whole video. Then same watermark data is utilized embed in all of the chosen frames to extend the probability of maintaining the watermark in manipulated watermarked video. For instance if some unauthorized person attempts to drop some frames of this very video, then if some watermarked frames dropped that are caused by the video, so if only one watermarked frame is left behind within the video then the watermark information might be recovered because of this frame only. The manipulations can easily be performed with video using either frame dropping or through some other way by any unauthorized person for illegal copying the clip. To save the overall quality of the video & have the algorithm more imperceptible the comprehensive video frame is not really altered by embedding the watermark information. Alternatively to the fact that the frame is split into blocks of 8 X 8 and applies singular values decomposition technique of these blocks. And watermark info is embedded during these singular values. Due to using several watermarked blocks, several watermarks might be recovered. So if any attack affects the watermarked image, many of the watermarks will survive. This block-by block method gives robustness against JPEG compression, cropping, blurring, Gaussian noise, resizing and rotation just like the results will reveal. The watermark can either be a pseudo-random number, or even a image. In this particular paper a picture is made use of as watermark. After watermarking the frames, we insert them in the recording over at their respective destinations to get the watermarked video. To extract the watermark from watermarked frames again same encryption secret is needed to obtain the watermarked frames. We put up key identifier to provide only 3 trials to the user. In the event the use tries extraction that have than 3 wrong keys then its assumed that they are attempting to find the watermarked frames by trying random keys. So at fourth try with wrong key the video is corrupted leaving no data behind[3].

## II. RELATED WORK

Doerr & Dugelay [4] have proposed video watermarking dependent on spread spectrum techniques in order to improve robustness. Here each watermark bit is spread over abundant chip rate (CR) and then modulated by way of a pseudo-random sequence of binary. This algorithms robustness increases with the increase of the variance of a given pseudo-noise sequence. Subsequently, the rise of (CR) will reduce the embedding rate of watermark information; on the other hand, the rise of variance may result in the perceptibility of one's watermark. The wavelet transform based video watermarking scheme was proposed by Liu et al [5] which treated embedding multiple information bits into the uncompressed video sequences. The embedding in LL sub-band made use of for reducing error probabilities of detection of BHC code A new type of watermarking scheme proposed by Niu et al [6] using two-dimensional and three – dimensional multi resolution signal decomposing. The watermark image that's decomposed with different resolution is embedded within the corresponding resolution of a given decomposed video. The robustness of watermarking is enhanced by coding the watermark information utilizing the Hamming error correction code. This approach is powerful against attacks an example would be frame dropping, averaging and lossy compression. A novel blind watermark algorithm based on SVD and DCT by Fen Lie et al [7] describes the fact that algorithm satisfies the transparence and robustness of this very watermarking system too. The experimental results demonstrate that this method is robust against common signal processing attacks. The digital video watermarking algorithm using Principal Component Analysis by Sanjana et al [9] proposed the invisible high bit rate watermark. Finally it was robust against various attacks for instance filtering, contrast adjustment, noise addition and geometric attacks. Haneih [8] have proposed a multiplicative video watermarking scheme with Semi-Blind maximum likelihood decoding for copyright protection.

### III. PROPOSED SYSTEM

**Singular value decomposition (SVD)**

Singular Value Decomposition (SVD) is mathematical technique for diagonal matrices in which the transformed domain involves basis states that are optimal. The singular value decomposition (SVD) is a trade in which you of representing a image inside a matrix for with most application in image processing. The singular value decomposition of causing complex matrix X is given by (2)

$$X = U\, SV^* \quad (2)$$

Where U is undoubtedly an m × m real or complex unitary matrix, S can be an m × n rectangular diagonal matrix with nonnegative real numbers according to the diagonal, and V* can be an n × n real or complex unitary matrix. The diagonal entries of S are known as the singular values regarding a and are also assumed to remain arranged in decreasing order the columns of a given U matrix are known as the left singular vectors while columns of the V matrix are named the very best singular vectors regarding a. Singular value of the matrix shows the luminance of an video frame layer while corresponding two of singular vectors specifies the geometry of this very video frame layer. Among the SVD-based watermarking, an video frame will be treated as a matrix, which further broke by SVD base method directly into three matrices such as U, S and V. subtle changes in the aspects of matrix S does not affect visual perception of the true quality of the cover video frame, SVD-based watermarking algorithms add the watermark information into the singular values of this very diagonal matrix S in such a way to meet the imperceptibility and robustness requirements of effective digital image watermarking algorithms.

In SVD based watermarking, proposed two effective, robust and imperceptible video watermarking algorithms. The 2 algorithms are based on the algebraic transform of Singular Value Decomposition (SVD). In the initial algorithm, watermark bit information are embedded among the SVD-transformed video within the diagonal-wise fashion, and then in the 2nd algorithm bits are embedded within the blocks-wise fashion. The diagonal-wise based algorithm achieved better robustness results, even though the block-wise algorithm gave higher data payload rate. Each algorithm embeds the watermark inside the transform-domain YCbCr space thus spreading the watermark in every frame of the video. The earliest algorithm suggests hiding watermark information within the diagonal-wise manner available as one of three SVD matrices; U, S and V. Nevertheless, the 2nd algorithm hides the watermark information within a block-wise manner in either the U or V matrices[1].

DWT-SVD video watermarking algorithm: The proposed DWT-SVD watermarking algorithms encompass two procedures, the very first embeds the watermark straight into the original video playback, even though the other extracts it make up the watermarked version of the clip.

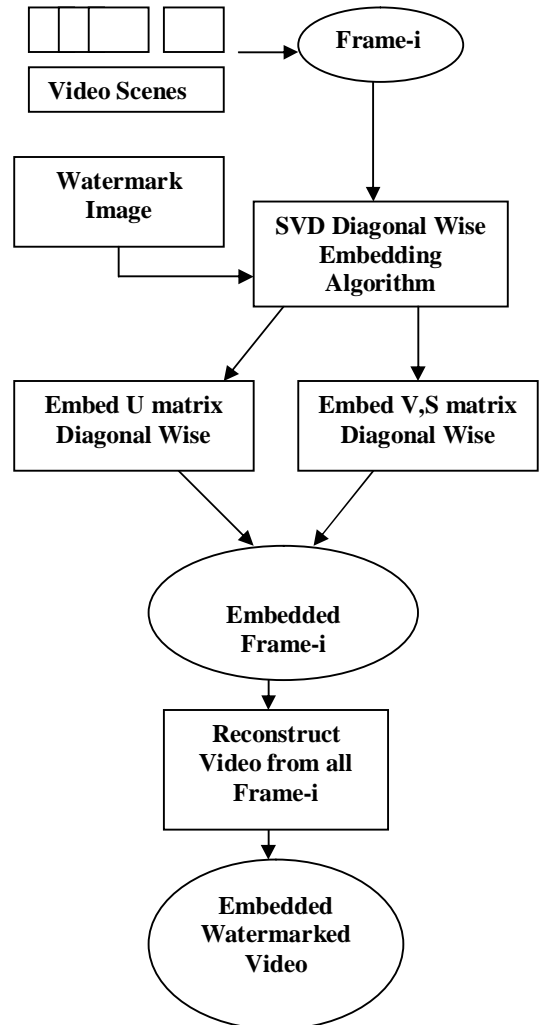

**Fig 1**. Embedding Diagonal Wise Video Watermark Process

**Steps In fig 1.**

1. Video scenes are extracted from the source video.
2. For each scene it will extract multiple frames.
3. For each frame in the scene randomly it will select the frame-i and then embedding process is activated.
4. In the embedding process watermark image along with frame-i is taken as input. In this process SVD algorithm extracts three matrices namely S,V,U for embedding purpose for embedding process.
5. Diagonal elements in the S,V,U matrices are extracted and embedding using watermark image.
6. Finally Video is constructed using embedded images .

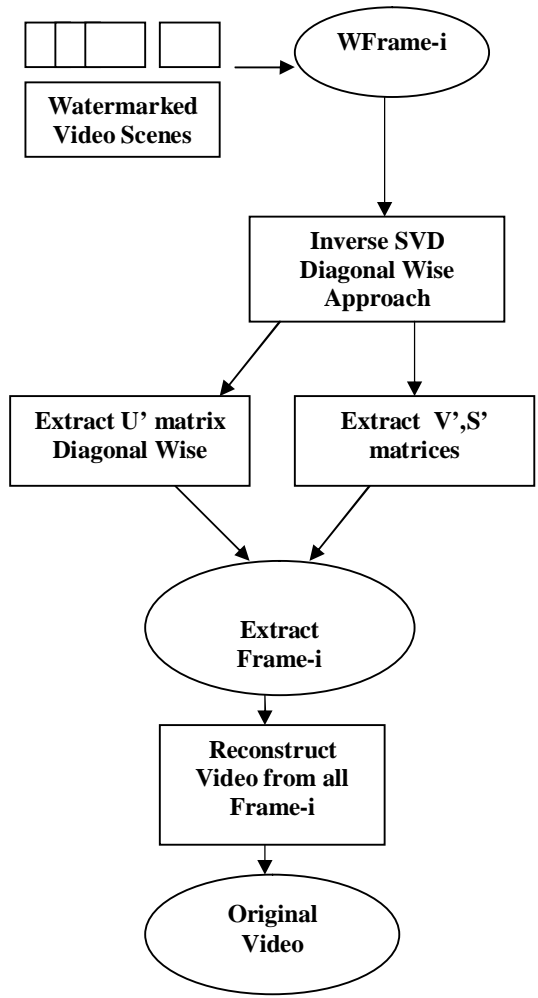

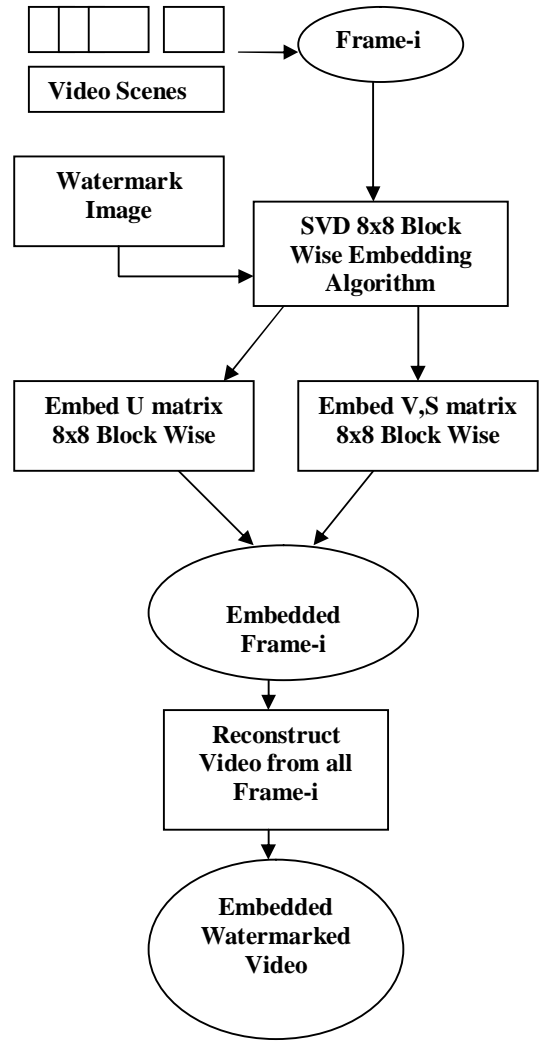

**Fig 2.** Watermark Extraction Process in Diagonal Wise Process:

**Steps In fig 2.**

7. For each watermarked scene in embedded video , extract watermark embedded frames.

8. For each embedded frame wframe-i is processed using Inverse SVD for extraction process.

9. Inverse SVD algorithm extracts three matrices namely S',V',U' .

10. Embedded Diagonal elements S',V',U' matrices are extracted from Wframe-i.

11. Finally Original Video is constructed using extracted frames .

**Fig 3.** Embedding Block Wise Video Watermark Process

**Steps In fig 3.**

12. Video scenes are extracted from the source video.
13. For each scene it will extract multiple frames.
14. For each frame in the scene randomly it will select the frame-i and then embedding process is activated.
15. In the embedding process watermark image along with frame-i is taken as input. In this process SVD algorithm extracts three matrices namely S,V,U for embedding purpose for embedding process.
16. 8x8 block wise elements are extracted from the S,V,U matrices and embedding using watermark image.
17. Finally Video is constructed using embedded images **.**

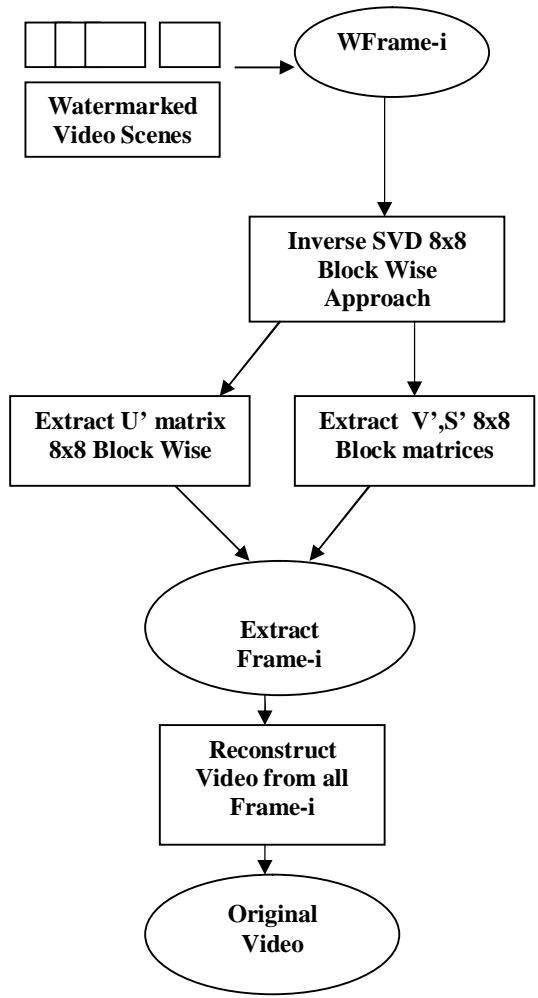

**Fig 4.** Watermark Extraction Process in Block Wise Process:

**Steps In fig 4.**

18. For each watermarked scene in embedded video , extract watermark embedded frames.
19. For each embedded frame wframe-i is processed using Inverse SVD for extraction process.
20. Inverse SVD algorithm extracts three matrices namely S',V',U' .
21. Embedded 8x8 block wise S',V',U' matrices are extracted from Wframe-i.
22. Finally   Original Video is constructed using extracted frames .

### IV RESULTS:

All experiments were performed with the configurations Intel(R) Core(TM)2 CPU 2.13GHz, 2 GB RAM, and the operating system platform is Microsoft Windows XP Professional (SP2).  This framework requires third party libraries like jama, jai,jmf.

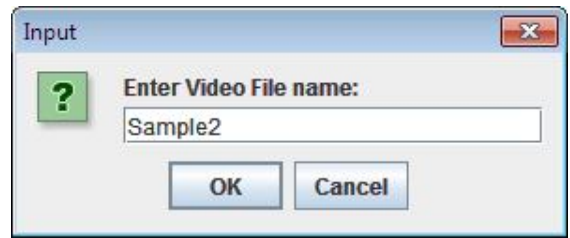

Fig 1: Loading Video File

Video duration: 69955000
No. times message repeated: 20
Message interval (long): 3331190
Writing output to Sample2Out.flv
...
Adding message at time 3333333
A =
  2.421649   2.151528   0.799923   0.935598   2.114568
  2.694066   2.013241   1.185589   1.202511   2.158039
  2.352369   1.894765   0.933314   0.983476   1.959936
  4.460927   3.692423   1.688805   1.830044   3.762587
  3.243859   2.869148   1.080296   1.261010   2.824390
  3.530655   2.750001   1.478701   1.506847   2.899877
  3.621949   3.160936   1.231963   1.439773   3.122544
  3.574548   3.093920   1.243018   1.418447   3.077635

LSB BITWISE OPERATIONS:
A = U S V^T

U =
  0.262971   0.302902  -0.194761   0.264837   0.058581
  0.280483  -0.578541   0.588850  -0.064835   0.030849
  0.249373  -0.173853  -0.158872   0.265674  -0.127749
  0.475946  -0.092292  -0.179463  -0.210128  -0.776014
  0.351815   0.374245  -0.128854  -0.668089   0.240748
  0.371435  -0.484825  -0.459566  -0.026636   0.529714
  0.391271   0.312253   0.574487  -0.013454   0.187631
  0.385530   0.249363   0.006065   0.603066   0.057769

S =
 15.423896   0.000000   0.000000   0.000000   0.000000
  0.000000   0.557828   0.000000   0.000000   0.000000
  0.000000   0.000000   0.028258   0.000000   0.000000
  0.000000   0.000000   0.000000   0.000000   0.000000
  0.000000   0.000000   0.000000   0.000000   0.000000
  0.000000   0.000000   0.000000   0.000000   0.000000

V =
  0.606212  -0.217286   0.083245  -0.554027   0.520978
  0.507058   0.566092   0.140119  -0.209500  -0.599090
  0.224974  -0.680685  -0.442008  -0.057397  -0.536087
  0.247221  -0.372042   0.777197   0.429180  -0.110615
  0.513487   0.174869  -0.417170   0.679466   0.264665

Correlation number = 4.973278107431424E16
2-norm = 15.423896392011654
singular values =
  15.423896   0.557828   0.028258   0.000000   0.000000
0

```
A =
   4.234966  3.110914  1.702273  3.008802  3.035072
   4.073014  2.989882  1.679930  3.127247  2.828290
   2.952330  2.170997  1.250097  2.376431  2.023642
   5.592918  4.109054  2.279944  4.120280  3.957800
   2.741695  2.021926  1.118196  1.926302  1.999619
   4.669681  3.439546  1.935295  3.483350  3.318866
   4.983467  3.644955  1.982448  3.639116  3.482471
   1.688690  1.242230  0.702901  1.294698  1.182177
```

LSB BITWISE OPERATIONS:
A = U S V^T

```
U =
   0.366984  -0.452411  -0.002915  -0.806006  -0.009048
   0.356906   0.397963   0.080917  -0.156644   0.308460
   0.261192   0.654652  -0.135630  -0.211511  -0.374186
   0.487398  -0.084186  -0.045522   0.272164   0.461803
   0.238247  -0.410632  -0.390770   0.287337  -0.470110
   0.408892   0.035662  -0.473600   0.221926   0.161519
   0.431533  -0.102350   0.766148   0.273497  -0.270854
   0.148288   0.144663  -0.094847  -0.019780  -0.480666

S =
  19.027182   0.000000   0.000000   0.000000   0.000000
   0.000000   0.291300   0.000000   0.000000   0.000000
   0.000000   0.000000   0.066064   0.000000   0.000000
   0.000000   0.000000   0.000000   0.000000   0.000000
   0.000000   0.000000   0.000000   0.000000   0.000000

V =
   0.602742  -0.199778   0.562832  -0.324232   0.418193
   0.442724  -0.148260   0.106528   0.857652  -0.187347
   0.245937   0.114977  -0.661430   0.125326   0.687821
   0.446460   0.824694  -0.041647  -0.150473  -0.310124
   0.425307  -0.494740  -0.482335  -0.347790  -0.469830
```

Correlation number = 8.324124881633576E16
2-norm = 19.027182027128866
singular values =
   19.027182   0.291300   0.066064   0.000000   0.000000

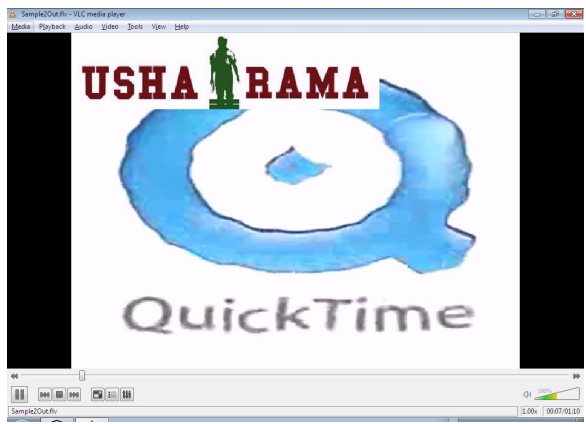

**Fig 2: Watermarked Video Using Proposed SVD Embedding Process**

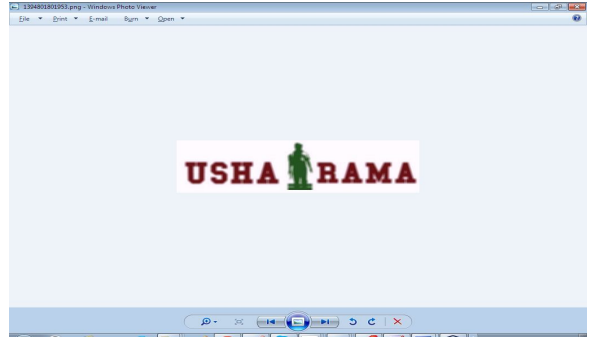

**Fig 3: Extracted Image from Video after Extraction Process**

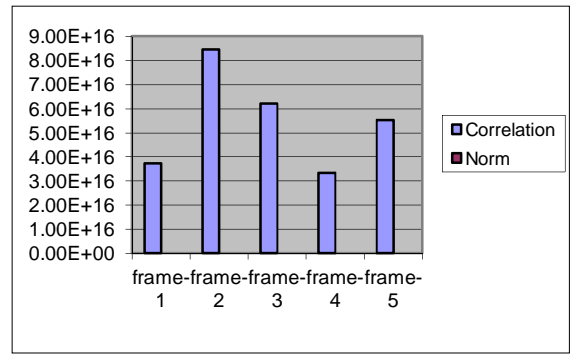

**Fig 4: Graph comparison between correlation with Norm of the S,V,U matrices**

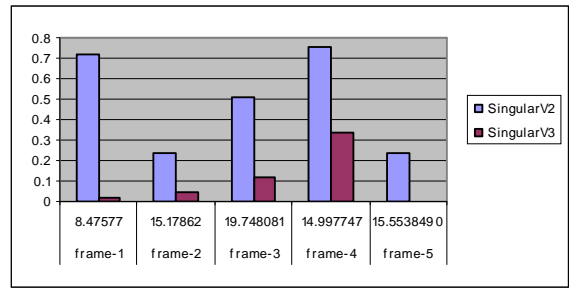

**Fig 5: Bar Graph Performance analysis between singular matrices in embedding process.**

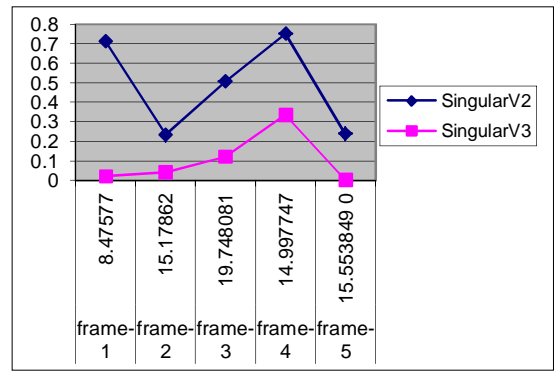

**Fig 5: Each Frame Performance analysis between singular matrices in embedding process.**

## V. CONCLUSION

Robustness of proposed methods is carried out by a variety of attacks. When the robustness values of three algorithms are compared, for all the attacks the DCT-SVD algorithm has given best robustness. The proposed algorithm results are compared with existing algorithms. Proposed approach performs well for video with specific size of embedded image. This system has some limitations 1. Due to the noise in the watermark or original video frames embedding and extraction process is problematic. 2. Proposed block or diagonal element based approaches are suitable to only specified standard video formats. 3. if the size of the video increases then frame size also increases then this framework may not suitable. In future these limitations can be overcome by proposing robust watermarking schemes.

## REFERENCES

[1] A Robust Video Watermark Embedding and Extraction Technique Based on Random Frame Selection Amrinder Singh, IJRIT International Journal of Research in Information Technology, Volume 2, Issue 2, February 2014, Pg: 28-37.

[2] Haneih Khalilian, Ivan V.Bajic, "Multiplicative Video watermarking with Semi-Blind Maximum Likelihood Decoding for Copyright Protection". IEEE Conference: 125-130, 2011.

[3] Sanjana Sinha, Prajnat Bardhan, Swarnali Pramanick, Ankul Jagatramka, Dipak K.Kole, Aruna Chakraborty, "Digital Video Watermarking using Discrete Wavelet Transform and Principal Component Analysis", International Journal of Wisdom Based Compuitng.Vol.1 (2): 7–12. 2011

[4] A Novel Block based Video in Video Watermarking Algorithm using Discrete Wavelet Transform and Singular Value Decomposition L.Agilandeeswari, International Journal of Advanced Research in Computer Science and Software Engineering.

[5] Video Watermarking Algorithms Using the SVD Transform Lama Rajab, European Journal of Scientific Research ISSN 1450-216X Vol.30 No.3 (2009), pp.389-401 © EuroJournals Publishing, Inc. 2009..

[6] Feng Liu, Ke Han, Chang zheng Wang, "A Novel Blind Watermark Algorithm based on SVD and DCT". IEEE Conference: 283–286, 2009.

[7] Doerr G and Dugelay J L, "A guide tour of video watermarking". Signal Processing: Image Communication 18(4): 263–282, 2003.

[8] Liu H, Chen N, Huang J, Huang X and Shi Y Q , "A robust DWT-based video watermarking algorithm".Proc. IEEE Int. Sym. Circuits and Systems, Scottsdale, Arizona 3: 631–634, 2002.

[9] Niu X, Sun S and Xiang W, "Multiresolution watermarking for video based on gray-level digital watermark" IEEE Transactions on Consumer Electronics 46(2): 375–384.2000